\documentclass[final,3p,times,twocolumn,authoryear]{elsarticle}

\usepackage{amssymb}
\usepackage{amsmath}
\usepackage{graphicx}%
\usepackage{multirow}%
\usepackage{amsmath,amssymb,amsfonts}%
\usepackage{amsthm}%
\usepackage{mathrsfs}%
\usepackage[title]{appendix}%
\usepackage{xcolor}%
\usepackage{textcomp}%
\usepackage{manyfoot}%
\usepackage{booktabs}%
\usepackage{algorithm}%
\usepackage{algorithmicx}%
\usepackage{algpseudocode}%
\usepackage{listings}%
\usepackage[version=4]{mhchem}
\usepackage{lineno}
\usepackage{longtable}
\usepackage{booktabs}
\usepackage{chemformula}
\usepackage{ltablex}
\keepXColumns
\usepackage{threeparttablex}
\usepackage{pdflscape}
\usepackage{threeparttable}
\usepackage{adjustbox}

\journal{Icarus}

\begin{document}

\begin{frontmatter}



\title{Novel Chemical Pathways for the Formation of Nucleobase Precursors via Benzene $\pi$-Bond Addition to HCN} 
\tnotetext[cor1]{Corresponding author}
\fntext[deceased]{Deceased 16 March 2026}
\author[label1,label2,label3]{Jeehyun Yang\corref{cor1}}
\ead{jeehyuny@uchicago.edu}
\author[label4]{Danica J. Adams}
\author[label2,label3]{Renyu Hu}
\author[label2,label3]{Yuk L. Yung\fnref{deceased}}
\affiliation[label1]{organization={Department of Astronomy and Astrophysics, The University of Chicago},
            addressline={5640 South Ellis Avenue},
            city={Chicago},
            postcode={60637},
            state={Illinois},
            country={USA}}

\affiliation[label2]{organization={Division of Geological and Planetary Sciences, California Institute of Technology},
            addressline={1200 E California Blvd.},
            city={Pasadena},
            postcode={91125},
            state={CA},
            country={USA}}

\affiliation[label3]{organization={Jet Propulsion Laboratory, California Institute of Technology},
            addressline={4800 Oak Grove Drive},
            city={Pasadena},
            postcode={91109},
            state={CA},
            country={USA}}

\affiliation[label4]{organization={Department of Earth and Planetary Sciences, Harvard University},
            addressline={20 Oxford Street},
            city={Cambridge},
            postcode={02138},
            state={MA},
            country={USA}}



\begin{abstract}
We propose a simple and efficient pathway for the formation of precursors to core nucleobases in DNA and RNA using a suite of computational chemistry methods. Benzene, which is thermochemically stable in \ce{N2}- or \ce{CO2}-dominated atmospheres, could have formed via upper-atmospheric photochemistry or surface lightning and accumulated on the early Earth or Mars. However, nitrogen insertion into the benzene ring to form pyrimidine and purine is widely considered to be challenging. We propose that nitrogen incorporation occurred through HCN 1,4-cycloaddition to benzene’s $\pi$-system, followed by a \ce{C2H2} fragmentation mechanism, as confirmed by quantum chemistry calculations. This pathway, potentially facilitated by photochemistry at the ocean surface or episodic impact events on local reservoirs, can lead to pyrimidine formation, which can further react with \ce{NH3} and \ce{HCN} to produce purine. Extending this pathway to early Mars, our photochemical model simulates heterocyclic compound formation under cold, dry surface conditions that favor high benzene and HCN concentrations but lack liquid water. We thus propose that organics formed during dry phases may have later dissolved into surface waters during wet phases and become concentrated as ocean sediments. This result supports Mars Sample Return efforts focused on ancient aqueous environments likely to retain prebiotic signatures.
\end{abstract}





\end{frontmatter}



\section{Introduction}\label{sec1} 

Since the pioneering experiments by Miller and Urey \citep{miller1953production}, which simulated early Earth conditions and synthesized organic compounds, including amino acids, from basic molecules such as \ce{H2O}, \ce{NH3}, \ce{CH4}, and \ce{H2}, many subsequent studies have built upon these so-called ``Miller-type'' experiments by varying the starting gases and energy sources, producing compelling experimental results that further support the prebiotic synthesis of amino acids in reducing atmospheres \citep{sanchez1966cyanoacetylene, oro1961synthesis, oro1962synthesis, civivs2004amino, cleaves2008reassessment, johnson2008miller, johnson2009diversity, parker2011primordial, horst2012formation, mccollom2013miller, wollrab2016chemical}. In the 1980s, ribonucleic acid (RNA) began gaining increasing attention in origin-of-life research alongside deoxyribonucleic acid (DNA), as nucleic acids play a fundamental role in genetic information storage and transmission in modern biology. In particular, the RNA World hypothesis proposes that RNA may have served both catalytic and informational roles in early stages of life’s emergence \citep{gilbert1986origin, joyce1989rna}. Consequently, another research stream has emerged, focusing on the synthesis of RNA and DNA. 

In modern biochemistry, genetic polymers such as RNA and DNA are composed of nucleotides, which consist of a nucleobase attached to a ribose or deoxyribose sugar and a phosphate group; without the phosphate group, the molecule is referred to as a nucleoside. However, the abiotic synthesis of complete nucleotides under plausible prebiotic conditions remains challenging, as coupling reactions between nucleobases, sugars, and phosphate groups are often inefficient or require specific environmental conditions \citep{powner2009synthesis, szostak2012eightfold, hud2013origin, sutherland2017opinion}. Alternative routes to nucleotide precursors have therefore been proposed \citep{powner2009synthesis,patel2015common, becker2019unified}. Consequently, many origin-of-life studies have focused on identifying plausible pathways for the formation of nucleobases and related heterocyclic precursors that could later participate in the assembly of nucleosides and nucleotides. 

Nucleobases are nitrogen-containing heterocycles that form the informational units of nucleic acids. The nucleobases used by life today are adenine (A), guanine (G), cytosine (C), thymine (T), and uracil (U). These nucleobases belong to two structural classes: \textit{pyrimidines}, which consist of a single six-membered heterocyclic ring (cytosine, thymine, and uracil), and \textit{purines}, which contain a fused bicyclic ring system composed of a five-membered ring fused to a six-membered ring (adenine and guanine). In this work, the terms pyrimidine (\ce{C5H5N}) and purine (\ce{C5H4N4}) refer to the corresponding parent heterocyclic structures that serve as precursors to these nucleobases. Although the formation of some nucleobases has been experimentally demonstrated in a Titan-like reducing atmosphere \citep{horst2012formation} or in a reducing atmosphere containing \ce{NH3} and CO, which highlights the role of formamide as an intermediate in nucleobase formation \citep{ferus2017formation}, the specific chemical pathways leading to their formation in anoxic but non-reducing atmospheres (e.g., \ce{N2}- or \ce{CO2}-dominated) remain unclear in detail. Such \ce{N2}- or \ce{CO2}-dominated atmospheres are less reducing, which may limit the efficient production of key precursors such as HCN while promoting competing oxidation pathways.

However, it is widely accepted that hydrogen cyanide (HCN) plays a crucial role in the prebiotic formation of nucleobases. HCN is directly and abundantly produced from various gas mixtures, including \ce{NH3}, by electrical discharge \citep{miller1955production, ferus2017formation} and photochemistry \citep{kaye1983hcn, zahnle1986photochemistry, ferris1988formation}, and is well known to polymerize to form adenine \citep{miller1957mechanism, oro1960synthesis}. Its miscibility with water also facilitates rapid and efficient polymerization into other complex structures \citep{ruiz2012new, bonnet2013compositional}, although HCN can also undergo hydrolysis in aqueous environments \citep{miyakawa2002cold}. In addition, HCN participates in a variety of prebiotically relevant chemical processes, including the formation of metal–cyanide complexes such as ferrocyanide \citep{todd2024favorable} and cyanosulfidic reaction networks that generate precursors to biomolecules \citep{patel2015common}. These unique properties highlight HCN's potentially pivotal role in the origin of life, making HCN oligomerization a subject of extensive investigation in both laboratory \citep{oro1961synthesis, oro1962synthesis} and theoretical studies \citep{oro1961mechanism, glaser2007adenine, roy2007chemical, benallou2016understanding, benallou2017mechanism, benallou2019new}. 

Despite HCN's promising role in the prebiotic formation of nucleobases, several considerations should be noted: (i) Previous prebiotic chemistry experiments often used HCN concentrations much higher than those likely available in prebiotic environments, making them less representative of early Earth's chemical conditions \citep{oro1960synthesis, oro1961mechanism, ferris1966unusual, sanchez1966cyanoacetylene, ferris1978hcn}. Given solar abundances, nitrogen atoms are approximately four times less abundant than carbon \citep{Lodders-2020}, and most nitrogen would have been present as \ce{N2} due to its thermochemical stability, a highly unreactive molecule compared to other nitrogen-bearing species such as \ce{NH3} and \ce{HCN}; (ii) HCN oligomerization requires multiple steps,  traversing at least four different potential energy surfaces \citep{glaser2007adenine, roy2007chemical, benallou2016understanding, benallou2017mechanism, benallou2019new}, making it highly susceptible to disruption by other competing reactions in natural environments (e.g., various volatiles, hazes, dissolved compounds in liquid phases, mineral surfaces). This complexity suggests that selectively guiding HCN molecules to react with one another in a prebiotic setting would be challenging; (iii) The previously proposed pathway first forms a five-membered ring structure, followed by ring closure to form a six-membered ring \citep{glaser2007adenine, roy2007chemical, benallou2016understanding, benallou2017mechanism, benallou2019new}. This pathway primarily accounts for the formation of adenine (A) and possibly guanine (G), which are classified as purine-type nucleobases. However, thymine (T), cytosine (C), and uracil (U) each contain only a single six-membered ring structure. Thus, the previously suggested HCN oligomerization pathway cannot explain the formation of these pyrimidine-type nucleobases, making it an unlikely general mechanism. More plausibly, pyrimidine (\ce{C4H4N2}) and purine (\ce{C5H4N4}) might form first, as these core structures can serve as precursors to all nucleobases \citep[A, G, T, C, and U;][]{oro1961synthesis}.

Thus, understanding the chemical pathways leading to the formation of these two core building blocks of life is crucial for bridging the gap between laboratory conditions that enable these reactions and the environments of early Earth. In this work, we employ a state-of-the-art rate-based automatic chemical reaction mechanism generator to investigate the potential precursors of nucleobases and use \textit{ab initio} calculations to propose novel chemical pathways for the formation of the two core building blocks of the origin of life: pyrimidine and purine. We further assess the broader plausibility of these pathways within the context of early Earth’s prebiotic environment and evaluate their feasibility under atmospheric conditions relevant to early Mars.

\section{Methods}\label{sec: methods}

\subsection{Automatic Chemical Network Generation for the nucleobases Thermal Dissociation}\label{sec: rmg}

To investigate the potential precursors of nucleobases, we adopted a reverse, decomposition-based approach. Rather than attempting to enumerate the many possible combinations of precursor molecules that could assemble into nucleobases, we instead examined the thermal decomposition of the five canonical nucleobases (A, G, T, C, and U) under conditions representative of primitive Hadean Earth. The rationale is that when complex molecules are subjected to energetic environments (in this case, thermal energy), they tend to break down into thermochemically stable fragments. These stable species are more likely to accumulate and therefore represent plausible building blocks for prebiotic chemistry. This approach therefore enables the identification of thermochemically stable fragments that may serve as plausible precursor molecules for prebiotic chemistry.

To systematically explore these decomposition pathways, we employed the automated chemical reaction network generator Reaction Mechanism Generator (\texttt{RMG}) \citep{Gao_2016, RMG-database, rmg-v3, RMG-developers}. \texttt{RMG} is a \texttt{python}-based open-source software that constructs chemical reaction networks by iteratively identifying kinetically relevant reactions using a rate-based algorithm, as described extensively in previous studies \citep{Gao_2016, rmg-v3, RMG-database}. Given initial chemical conditions ($T$, $P$, and species mixing ratios), \texttt{RMG} systematically enumerates possible reaction pathways using its reaction families and thermochemical database. This automated approach reduces human bias and enables objective and comprehensive generation of chemical reaction networks. 

In the current work, temperatures ranging from 300 to 750 K and pressures between 10$^{-7}$ and 10$^{2}$ bar are chosen to encompass a range of early earth atmospheric conditions, from the period following the solidification of the initial magma ocean \citep{elkins2008linked} to the late Hadean era, when liquid water may have been present on the surface \citep{wilde2001evidence, sleep2001initiation, valley2002cool}. The initial molecular mixing ratio was set to 20 \% for each adenine, guanine, cytosine, thymine, and uracil. This choice is not intended to represent realistic prebiotic conditions but rather ensures that each nucleobase is present in sufficient quantity for the reaction network generator to explore its decomposition pathways and identify thermochemically stable fragments. The reaction time was set to 0.6 billion years (long enough to reach thermal equilibrium), reflecting the timescale of prebiotic chemistry. Although the precise timing of the emergence of life remains uncertain, it is generally thought to have taken place sometime between the solidification of the magma ocean ($\sim$4.3 Ga) and the earliest geochemical evidence for life around $\sim$3.7 Ga \citep{wilde2001evidence, valley2002cool}. After the model generation was completed, the final model contained 151 species and 879 reactions, which can be found in \texttt{CHEMKIN} format in the Supplementary information, along with the \texttt{RMG} input file detailing the choice of the reaction and thermochemical libraries used to generate the reaction mechanism.

The reaction network generation approach employed in this work is used to explore the thermodynamically favored chemical space under conditions relevant to the thermal degradation of nucleobases. We emphasize that this approach is not intended to directly identify pathways leading to the formation of nucleobases, but rather to determine which chemical species are stable and likely to persist, and thus may subsequently participate in nucleobase synthesis. As such, thermodynamic stability does not necessarily imply kinetic accessibility, and additional analysis is required to evaluate specific reaction mechanisms. We further note that the full generated network (879 reactions) is not analyzed on a reaction-by-reaction basis; instead, it is used to identify dominant and persistent species arising from the thermal dissociation of nucleobases, which then motivate targeted investigation of specific reaction pathways.

\subsection{Thermochemical Equilibrium State Analysis}\label{sec: cantera}
Using the chemical network generated by \texttt{RMG}, we explored thermochemical equilibrium states along temperature$-$pressure profiles and assessed species stability. This was achieved by computing the thermochemical equilibrium state at various temperatures and pressures for a given initial molecular mixing ratio using \texttt{Cantera}, an open-source software written in \texttt{C++} \citep{Goodwin_Cantera_An_Object-oriented_2024}. Given geological evidence for liquid water on early Earth's surface \citep{wilde2001evidence, sleep2001initiation, valley2002cool}, we assumed the water vapor line (blue line in Figure~~\ref{fig:Fig1}b) as the upper boundary for the early Earth's temperature-pressure ($T$–$P$) profile. Under this assumption, we considered several background atmospheric compositions, including \ce{H2}-, \ce{H2O}-, \ce{N2}-, and \ce{CO2}-dominated atmospheres. For each atmospheric composition, benzene was introduced as a trace species with initial mixing ratios ranging from 1 ppt to 1\% (specifically 1 ppt, 1 ppb, 1 ppm, and 1\%). Each case was then allowed to reach thermal equilibrium along the water vapor $T$–$P$ profile. The result is shown in Figure~\ref{fig:Fig1}c.

\subsection{\textit{Ab Initio} Quantum Chemistry Calculations on the Gas-phase Potential Energy Surfaces (PES)}\label{sec: gaussian 09}
To investigate and validate the newly proposed chemical pathways leading to the formation of pyrimidine and purine, we performed \textit{Ab initio} quantum chemistry calculations for all stable species and transition states in this study. These calculations were conducted at the CBS-QB3 level of theory \citep{Montgomery-1999, Montgomery-2000} using \texttt{Gaussian 09} \citep{g09} and verified through intrinsic reaction coordinate (IRC) analyses \citep{Deng-1993-IRC, Deng-1994-IRC} at the B3LYP/CBSB7 level of theory \citep{g09}. All molecular parameter outputs are provided in the Supplementary Information.

\subsection{\textit{Ab Initio} Quantum Chemistry Calculations on the Liquid-phase Potential Energy Surfaces (PES)}\label{sec: cosmors}
To assess the effect of water solvation on the potential energy surface (PES) described in Section \ref{sec: gaussian 09}, we calculated kinetic solvent effects using a hybrid quantum chemistry and \texttt{COSMO-RS} approach, as described in Chung \textit{et al.,} 2023 \citep{chung2023solvent}. Briefly describing, geometries of reactants and transition states previously calculated at CBS-QB3 were re-optimized at the BP86/def2-TZVP level of theory using \texttt{TURBOMOLE 7.5} \citep{turbomole, balasubramani2020turbomole}. Single-point energy calculations were then performed in a virtual conductor using the \texttt{COSMO} solvation model at the BP86/def2-TZVPD level. These results were used to compute solvation free energies ($\Delta G_{\rm solv}$) at 298 K using \texttt{COSMOtherm (COSMO-RS)} \citep{sinnecker2006calculation}.

\subsection{Calculation of Rate-coefficients for Newly Proposed Chemical Pathways}\label{sec: arkane}

After completing the gas-phase PES calculations described in Section \ref{sec: gaussian 09}, we used \texttt{Arkane} \citep{Allen_2012, dana2023_arkane} to calculate temperature- and pressure-dependent rate coefficients, $k$($T,P$). To account for photochemical reactions (specifically, the photoexcitation of benzene and pyridine followed by their reactions with HCN), we adopted the photoexcitation chemistry framework described in \citet{yang2023high}. Briefly describing, we made several assumptions regarding the calculation of photoexcitation (i.e., S$_0 \rightarrow$ S$_1$) and phosphorescence (i.e., T$_1 \rightarrow$ S$_0$, involving a spin change) rate coefficients. These assumptions include the following: (i) 80\% of benzene molecules ($\Phi_{\rm ISC}$ = 0.8, adopted from \citet{duncan1981photoionization}) and 50\% of pyridine molecules ($\Phi_{\rm ISC}$ = 0.5, adopted from \citet{knight1976radiative}) in the ground state (S$_0$) immediately populate the lowest triplet state (T$_1$) following UV photoexcitation into the spin- and dipole-allowed first excited singlet state (S$_1$); (ii) the phosphorescence rates of T$_1$ benzene and pyridine are taken as the inverse of their respective lifetimes ($\tau$): 470 ns for benzene \citep{duncan1981photoionization} and 1 $\mu$s for pyridine \citep{terazima1988quantum}; and (iii) entrance barriers from those triplet states (instead of the ground states) are barrier-less reaction (i.e., reaction barrier is 0 kcal/mol). All calculated rate coefficients are provided in \texttt{CHEMKIN} format as supplementary information.

The photoexcitation rate ($J$) for benzene and pyridine was calculated in \texttt{KINETICS} using the following equation as a function of altitude (\textit{z}) and wavelength ($\lambda$):
 \begin{equation}
 J_{i}(\lambda,z) =
 \Phi_{i}(\lambda)\sigma_{i}(\lambda)F_0(\lambda)e^{-\sum \sigma_{j}(\lambda)n_jz}
 \end{equation}
 where $i$ denotes the target species (i.e., benzene or pyridine). $\Phi_i(\lambda)$ is the quantum yield of the photoexcitation reaction, taken as 0.8 for benzene \citep{duncan1981photoionization} and 0.5 for pyridine \citep{knight1976radiative}. $\sigma_i(\lambda)$ is the photoabsorption cross section of the target species, with values for benzene taken from \citet{Dawas-2017-benzene-cross-section} and for pyridine from \citet{bolovinos1984absolute}. $F_0(\lambda)$ represents the stellar photon flux at the top of the atmosphere (i.e., 250 km altitude) and was scaled to account for the faint young Sun (70\% of the current solar flux) and the 1.524 AU average distance between Mars and the Sun. The exponential term accounts for optical attenuation due to absorption by other atmospheric species, where $n_j$ is the number density of species $j$, and $\sigma_j(\lambda)$ is its UV absorption cross section. The vertical mixing ratio and temperature-pressure ($T$-$P$) profiles for these species are shown in Figure~ \ref{fig: mars}. As shown in Figure \ref{photons}, UV photons with wavelengths $\lambda \leq 236$ nm are depleted due to absorption by major abundant gas species. However, photons in the range of 236–300 nm still possess sufficient energy to drive barrierless reactions between benzene and HCN, and between pyridine and HCN. These reactions can proceed via well-skipping pathways, leading to the formation of pyridine + \ce{C2H2} and pyrimidine + \ce{C2H2}, respectively, as illustrated in Figure \ref{fig: fig_pyrimidine}.

\subsection{1D Photochemical Kinetic-Transport Atmospheric Modeling}\label{sec: KINETICS}
Unlike Earth, Mars lacks plate tectonics and therefore likely preserves ancient sedimentary records. With the Mars Sample Return mission scheduled for launch in 2030, it is critical to evaluate whether the early Martian atmosphere could have supported heterogeneous aromatic species formation. To quantitatively assess this possibility, we performed one-dimensional photochemical-kinetic transport modeling to determine the vertical mixing ratios and corresponding deposition rates of key heterocyclic aromatic species (pyridine and pyrimidine) in the early Martian atmosphere. In brief, we incorporated the heterogeneous aromatic species formation chemistry proposed in this study into the early Martian atmospheric model for the cold, dry epoch described by \citet{adams2025episodic}. We use the Caltech/JPL 1D Photochemical and Transport model named \texttt{KINETICS} \citep{allen1981vertical}. Previous works show early Mars may have experienced a CO-runaway state during dry, cold epochs \citep{zahnle2008photochemical,adams2025episodic}. We adapt a Mars \texttt{KINETICS} model based on \citet{nair1994photochemical} to the results of \citet{adams2025episodic}, which invoked a sink of atmospheric oxygen to the surface where it would oxidize reduced iron in the early crust. This drives a CO-runaway due to the depletion of atmospheric OH, which in turn cannot recycle CO back to \ce{CO2} efficiently. We consider their 1 bar atmosphere with the \ce{O2} sink of $10^9$ [molecules/cm$^2$/s], which corresponds to 7 meters of oxidation over approximately 600 Myr (the duration of the Noachian). We scaled the solar flux to reflect the faint young Sun, using 70\% of the present-day solar flux. We fix the CO, \ce{CO2}, \ce{N2}, and \ce{H2O} as well as the temperature-pressure profile according to their results. Our model solves the 1D continuity equation for the following species: O, O($^1$D), \ce{O2}, \ce{O3}, H, \ce{H2}, OH, \ce{HO2}, \ce{H2O2}, N, N($^2$D), NO, \ce{NO2}, \ce{NO3}, \ce{N2O}, \ce{N2O5}, \ce{HNO}, NH, \ce{NH2}, \ce{NH3}, \ce{N2H2}, C, CH, $^1$\ce{CH2}, \ce{CH2}, \ce{CH3}, \ce{CH4}, \ce{C2}, \ce{C2H}, \ce{C2H2}, \ce{C2H3}, \ce{C2H4}, \ce{C2H5}, \ce{C2H6}, \ce{C3H2}, \ce{C3H3}, \ce{C4H2}, \ce{C6H6}, \ce{HCO}, \ce{H2CO}, \ce{CH3CHO}, CN, HCN, HNC, \ce{H2CN}, CHCN, \ce{C2N}, \ce{HC2N2}, \ce{C3N}, \ce{H2C3N}, \ce{CH2NH}, NCO, HNCO, O$^+$, \ce{O2}$^+$, \ce{CO2}$^+$, \ce{CO2H}$^+$, C$^+$, CO$^+$, \ce{C5H5N} (pyridine), $^3$\ce{C5H5N} (triplet pyridine), \ce{C7H7N} (i1), $^3$\ce{C6H6} (triplet benzene), four \ce{C6H6N2} isomers (i2-1, i2-2, i2-3, and i2-4), three \ce{C4H4N2} isomers (pyridzaine, pyrimidine, and pyrazine).  We assumed a dry deposition velocity of 0.2 cm/s for heterocyclic aromatic species based on an analogy to benzene \citep{cohen-1994} due to their aromatic characteristics.

Beyond the C-H-O chemistry in \citet{nair1994photochemical}, the hydrocarbon network is based on that in \citet{willacy2022vertical} but with fewer species to simplify for the more oxidized environment compared to Titan's atmosphere. The nitrile network is based on both \citet{willacy2022vertical} and \citet{adams2021nitrogen}. We allow for H and \ce{H2} escape at the upper boundary conditions, fixing a velocity according to \citet{hunten1973escape}. We fix a mixing ratio of one percent \ce{CH4} at the lower boundary to source the hydrocarbon chemistry. We posit that this methane could be sourced from water-rock interactions or release from ice clathrates after obliquity changes \citep{etiope2013low, wordsworth2017transient, kite2020methane}. We also include a flux of benzene and HCN at the lower boundaries, sourced by lightning. Both HCN and benzene may be generated in high-energy environments such as lightning discharges in planetary atmospheres. Laboratory experiments simulating lightning chemistry have demonstrated efficient production of HCN from \ce{N2}-containing gas mixtures \citep{ferus2017high}. The formation of aromatic hydrocarbons such as benzene, however, depends sensitively on atmospheric composition, particularly the availability of reduced carbon species that can lead to acetylene and other hydrocarbon intermediates. Lightning would heat parcels of the atmosphere to thousands of Kelvin, temporarily allowing chemical equilibrium to describe the chemistry. We use \texttt{CEA} (Chemical Equilibrium with Applications, \citep{gordon1994computer}) to estimate the mixing ratios of benzene and HCN in these parcels. The time-averaged flux of these species from lightning then also depends on the lightning flash rate which scales with the convective available potential energy of the atmosphere. We adapt these parameters from \citet{adams2021nitrogen}. The new chemical pathways proposed in this work have also been implemented in \texttt{KINETICS}. Details of the thermochemical and photochemical rate‐coefficient calculations are provided in Section \ref{sec: arkane}.

\section{Results}\label{results}

\subsection{Thermochemical Stability of Benzene (\ce{C6H6}) in Various Atmospheric Scenario}\label{sec:benzene}

\begin{figure*}[hbt!]
\centering
\includegraphics[width=0.95\textwidth]{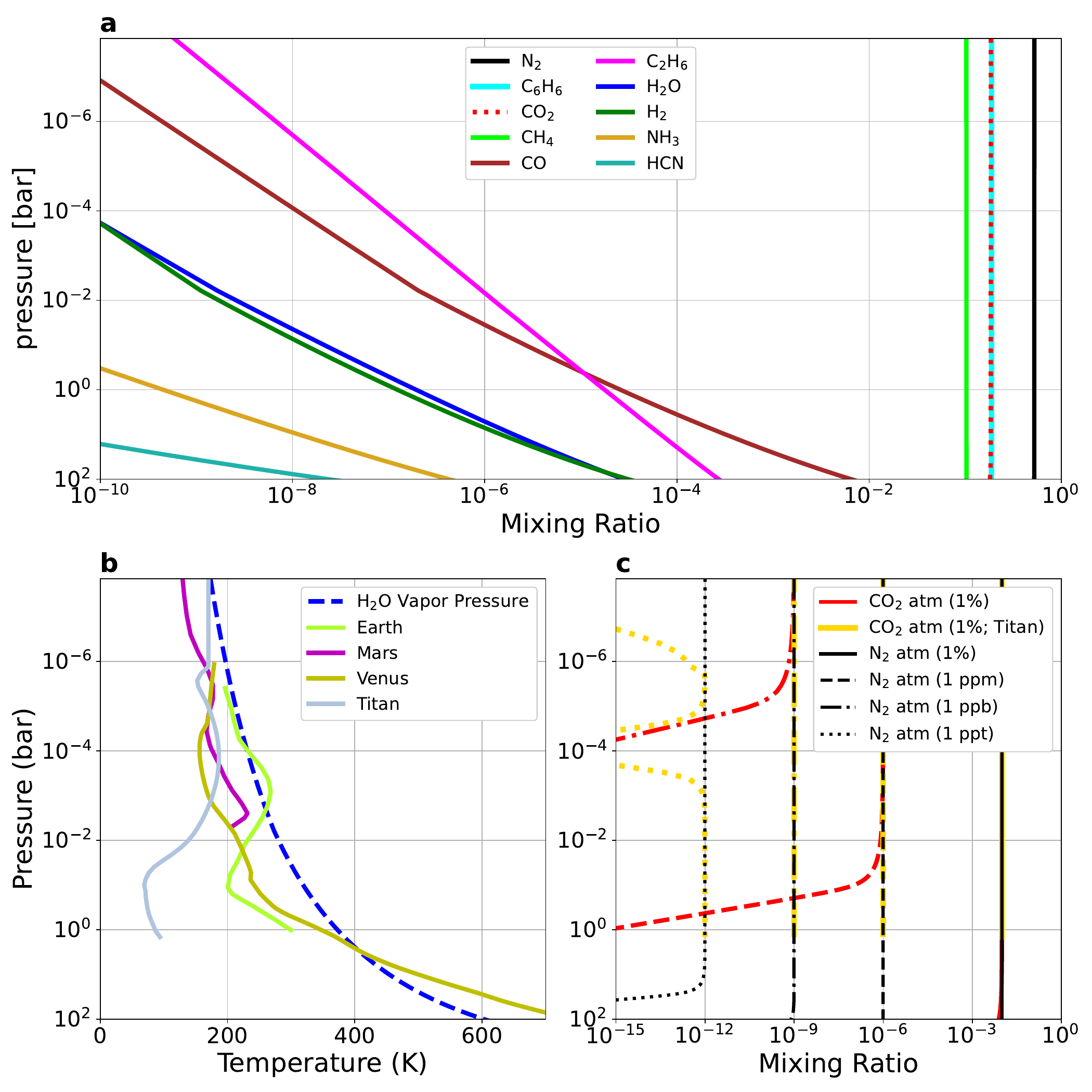}
\caption{\textbf{Thermochemical stability of various precursor species constituting nucleobases.} \textbf{(a)} Major species predicted for a thermally equilibrated mixture of nucleobases (20\% each of A, G, T, C, and U) at temperature and pressure conditions along the water vapor line (blue dashed line in panel\textbf{b}); \textbf{(b)} Various temperature–pressure profiles: \ce{H2O} vapor pressure line from \cite{Buck-1981}; Earth from \cite{hedin1987cospar}; Mars from the Mars Climate Database v6.1 \cite{forget1999improved, millour2019latest}; and Venus from \cite{bierson2020chemical}; Titan from \cite{willacy2022vertical}; \textbf{(c)} Thermochemical stability of benzene (\ce{C6H6}) along temperature–pressure conditions shown in panel \textbf{b} under different atmospheric compositions: \ce{CO2}-dominated (red), a \ce{CO2}-dominated Titan-like scenario (yellow), and \ce{N2}-dominated (black). Most curves follow the water-vapor temperature–pressure profile (blue dashed line in panel \textbf{b}), while the yellow lines correspond to a \ce{CO2}-dominated atmosphere whose $T$-$P$ structure follows the Titan profile (solid grey line in panel \textbf{b}). Numbers in parentheses indicate the initial benzene mixing ratio introduced before thermal equilibration. Solid, dashed, dash-dotted, and dotted lines correspond to initial benzene mixing ratios of 1\%, 1 ppm, 1 ppb, and 1 ppt, respectively. Note that the solid lines overlap each other. Although not shown, we also evaluated \ce{H2}- and \ce{H2O}-dominated atmospheres. Even when assuming an initial benzene mixing ratio of 1\%, the resulting benzene abundance at thermal equilibrium remains below 10$^{-23}$ when following the water-vapor temperature–pressure profile, indicating that benzene is thermochemically unstable in \ce{H2}- or \ce{H2O}-dominated atmospheres.}\label{fig:Fig1}
\end{figure*}

As shown in Figure~ \ref{fig:Fig1}a,  \ce{N2}, \ce{CO2}, \ce{CH4}, and \ce{C6H6} appear to be the dominant species under all $T$-$P$ conditions, each maintaining an abundance of at least 10\% abundance after thermal equilibrium. Among these, \ce{N2}, \ce{CO2}, and \ce{CH4} are widely recognized as primordial atmospheric gases on early Earth. Volcanic outgassing predominantly contributed oxidized species such as \ce{CO2} and \ce{N2} \citep{holland2020chemical, catling2017atmospheric}, while \ce{CH4} likely formed transiently through the reduction of \ce{CO} or \ce{CO2} via reactions with \ce{H2} produced by large asteroid impacts \citep{urey1952early, zahnle2020creation}.

Interestingly, \ce{C6H6} (benzene) was found to be a dominant chemical species under these conditions, suggesting its potential role as a precursor for nucleobase formation. Benzene is the first aromatic ring-structured molecule and has been extensively studied through both theoretical and experimental approaches. It is generally accepted in chemical kinetics that benzene formation primarily proceeds via the hydrogen-abstraction–acetylene-addition (HACA) mechanism starting from acetylene \citep{bittner1981composition, frenklach1985detailed}, with reaction rates precisely determined through combined modeling and experimental studies \citep{smith2020direct}. Observationally, benzene has been detected in planetary atmospheres, including those of Jupiter \citep{kim1985infrared}, Saturn \citep{bezard2001benzene}, Titan \citep{COUSTENIS2003383}, and possibly Mars \citep{freissinet2015organic, eigenbrode2018organic}. Its presence extends beyond planetary atmospheres to carbonaceous chondrites \citep{studier1965organic}, protoplanetary nebulae \citep{cernicharo2001infrared}, and the interstellar medium \citep{cooke2020benzonitrile}.

To further investigate benzene's stability, we evaluated its thermochemical equilibrium under various atmospheric conditions (Figure~ \ref{fig:Fig1}c). As expected, benzene is unstable in both reducing \ce{H2}-dominated and oxidizing \ce{H2O}-dominated atmospheres. Even at an initial mixing ratio of 1\% of benzene introduced, benzene remains chemically negligible in these environments (benzene mixing ratio less than $10^{-23}$). This instability arises because a \ce{H2}-dominated atmosphere can readily hydrogenate benzene, cracking it down to \ce{CH4}, while a \ce{H2O}-dominated atmosphere oxidizes benzene to \ce{CO} and \ce{CO2}. As a result, benzene is thermochemically unstable in \ce{H2}- or \ce{H2O}-dominated atmospheres.

It has to be noted that benzene detection in Jupiter and Saturn's atmospheres is likely attributed to photochemistry \citep{moses2000photochemistry} or ion-molecular chemistry \citep{wong2003benzene}, both of which require high-energy sources beyond thermal equilibrium. In deeper atmospheric layers (i.e., at a pressure exceeding 10 bar), these disequilibrium processes become insignificant \citep{visscher2010deep}. Instead, benzene is thermochemically converted to methane, which may later be transported back to the upper atmosphere, where photochemistry and ion chemistry regenerate benzene. The benzene abundance and stability in the deep part of the atmosphere would be crucial if benzene indeed takes an important role in the formation of nucleobases. Thus, \ce{H2}- and \ce{H2O}-dominated atmosphere might not be favorable for prebiotic formation of benzene and its subsequent role in nucleobase formation.

In \ce{CO2}-dominated atmospheres, \ce{CO2} dissociation produces oxidizing O atoms, but this oxidation pathway is less efficient and requires higher temperatures than oxidation driven by reactive species generated from \ce{H2O} dissociation (e.g., OH and O). As shown in Figure~ \ref{fig:Fig1}c, benzene remains abundant down to $10^{-2}$ bar when its initial concentration is 1 ppm. However, \ce{CO2} photolysis in the upper atmosphere generates oxygen atoms that facilitate benzene oxidation to \ce{CO} even at low temperatures \citep{yung1999photochemistry}. Thus, sustaining significant benzene levels requires a cooler $T$-$P$ profile, similar to those of Mars or Titan (indicated as purple and grey lines, respectively, in Figure~ \ref{fig:Fig1}b). Indeed, when we tested benzene stability under a Titan-like $T$-$P$ profile but with a \ce{CO2}-dominated atmosphere instead of \ce{N2}, benzene remained thermochemically stable even at ppt levels at the surface level of Titan (see dotted yellow lines in Figure~ \ref{fig:Fig1}c). This is consistent with the detection of benzene proxies on Mars \citep{freissinet2015organic, eigenbrode2018organic}. 

In \ce{N2}-dominated atmospheres, benzene exhibits even greater stability, as shown by the black lines in Figure~ \ref{fig:Fig1}c, with concentration remaining at 1 ppt ($10^{-12}$) down to 10 bar. Benzene's stability in \ce{N2}-dominated atmospheres aligns with the detection of benzene in Titan atmospheres \citep{COUSTENIS2003383}. The stability is primarily due to the strong inertness of \ce{N2}, whose triple bond requires extremely high temperatures and pressures, such as those found in combustion engines, or high-energy UV radiation and energetic particles, which are only available in the upper atmosphere. From this, we conclude that \ce{N2}-dominated atmospheres are the most favorable for benzene accumulation originally formed from upper atmospheric photochemistry \citep{moses2000photochemistry} or ion-chemistry \citep{wong2003benzene} onto planetary surfaces, while \ce{CO2}-dominated atmospheres also provide favorable conditions for benzene accumulation. Under early Earth-like $T$-$P$ conditions, benzene formed in the upper atmosphere could accumulate on the surface in both \ce{N2}- or \ce{CO2}-dominated atmospheres. Aromatic hydrocarbons, including benzene, produced photochemically can condense or adsorb onto organic aerosol particles, which subsequently settle through gravitational sedimentation or are removed by precipitation, allowing these compounds to be deposited onto the surface. In contrast, \ce{H2}- and \ce{H2O}-dominated atmospheres are unfavorable for benzene stability.

Taken together, these results indicate that aromatic hydrocarbons, particularly benzene, can emerge as robust and persistent intermediates under a range of prebiotic conditions, especially in \ce{N2}- and \ce{CO2}-dominated atmospheres. This suggests that benzene may serve as an important building block for subsequent chemical evolution toward more complex organic species.

Motivated by this finding, we next investigate reactions involving benzene and hydrogen cyanide, a key nitrogen-bearing species in prebiotic chemistry. In particular, we explore a representative 1,4-cycloaddition and fragmentation pathway using quantum chemical calculations to assess its kinetic feasibility.

\subsection{Nitrogen Substitutions in Benzene (\ce{C6H6}) and Pyridine (\ce{C5H5N}) via 1,4-HCN-cycloaddition/\ce{C2H2}-fragmentation Reactions}\label{sec:pyridine}
Given benzene's stability across a wide range of \ce{N2}-dominated atmospheres and certain \ce{CO2}-dominated conditions, a key question now is how nitrogen atoms can be incorporated into ring structures, as pyrimidine (\ce{C5H5N}) and purine (\ce{C5H4N4}) contain two and four nitrogen atoms, respectively. We propose nitrogen substitution in the benzene ring via HCN addition to the $\pi$-bond, a process known as 1,4-cycloaddition and fragmentation (CAF) \citep{comandini2011theoretical}. Before discussing the detailed reaction pathway, we briefly comment on the expected reactivity of the system. Unlike classical 1,4-cycloaddition reactions involving highly reactive or strained dienophiles, the present system involves HCN as a comparatively less reactive species. As such, this reaction would not be expected to proceed efficiently under purely thermal, ground-state conditions due to the aromatic stability of benzene. However, under activated conditions, such as photoexcitation or radical formation, the effective reactivity of benzene can be significantly enhanced. Under such conditions, addition pathways involving small molecules such as HCN may become kinetically accessible, as supported by the quantum chemical calculations presented below.

As shown in Figure~\ref{fig: 1st_insertion_ball_sticks}, this reaction begins with a pericyclic cycloaddition between two distinct $\pi$-electron systems, forming a ring with two new $\sigma$-bonds and two fewer $\pi$-bonds, analogous to the Diels-Alder reaction \citep{diels1928synthesen}. The subsequent fragmentation of acetylene (\ce{C2H2}) restores the newly formed $\sigma$-bonds to $\pi$-bonds, preserving the aromatic structure while introducing new substituents \citep{comandini2011theoretical}. The 1,4-CAF reaction between benzene (\ce{C6H6}) and ortho-benzyne (o-\ce{C6H4}) has been studied both theoretically \citep{comandini2011theoretical} and experimentally \citep{yang2022experiment}. Notably, this reaction involves the ortho-benzyne radical species adding to the $\pi$-bond site of a neutral benzene molecule, directly forming a fused ring structure without requiring multiple steps of hydrogen abstraction or elimination from parent molecules. As a result, the 1,4-CAF reaction has recently been proposed as an efficient pathway for the formation of larger molecules \citep{comandini2017polycyclic}. 

\begin{figure*}[hbt!]
\centering
\includegraphics[width=1\textwidth]{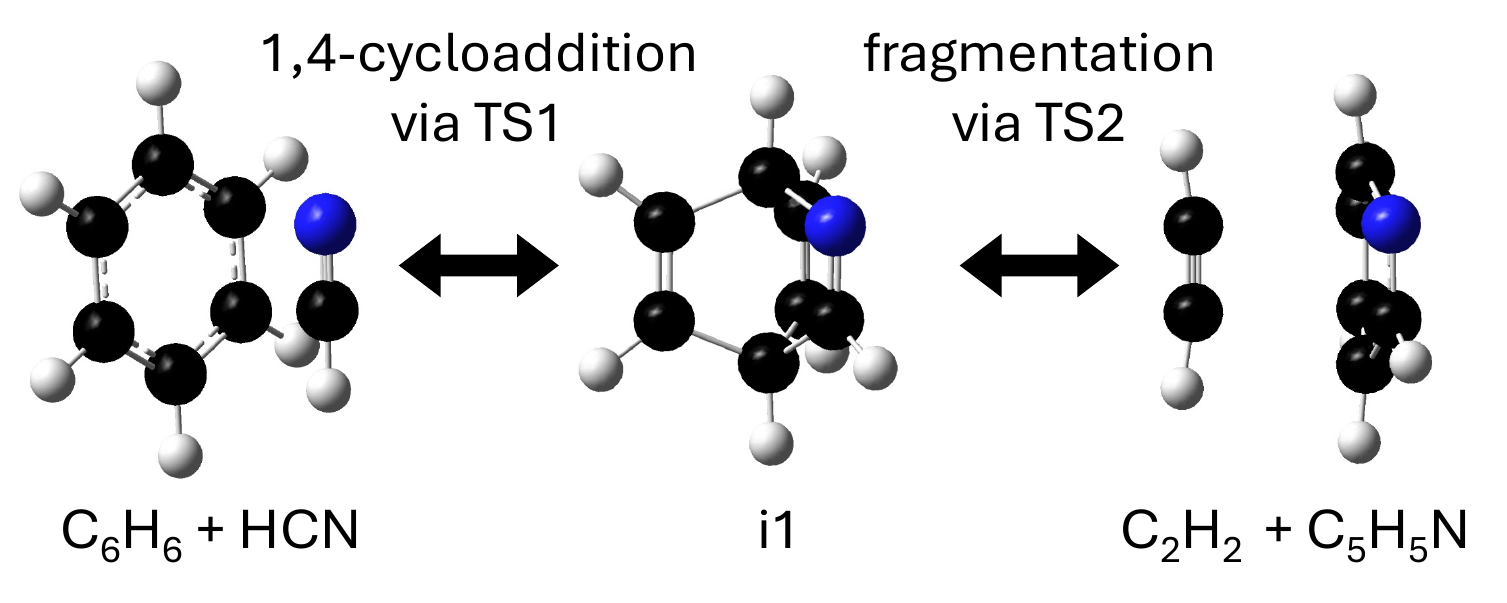}
\caption{Visualization of the reaction sequence for the 1,4-cycloaddition/fragmentation pathway, illustrated using the reaction between \ce{C6H6} and HCN as an example. The reaction proceeds through a 1,4-cycloaddition between \ce{C6H6} and HCN via transition state TS1, forming the intermediate i1, followed by fragmentation through transition state TS2 to yield \ce{C2H2} and pyridine (\ce{C5H5N}). Molecular structures are shown in ball-and-stick representation for clarity. Carbon atoms are shown in black, hydrogen in white, and nitrogen in blue.}\label{fig: 1st_insertion_ball_sticks}
\end{figure*}

\begin{figure*}[hbt!]
\centering
\includegraphics[width=0.95\textwidth]{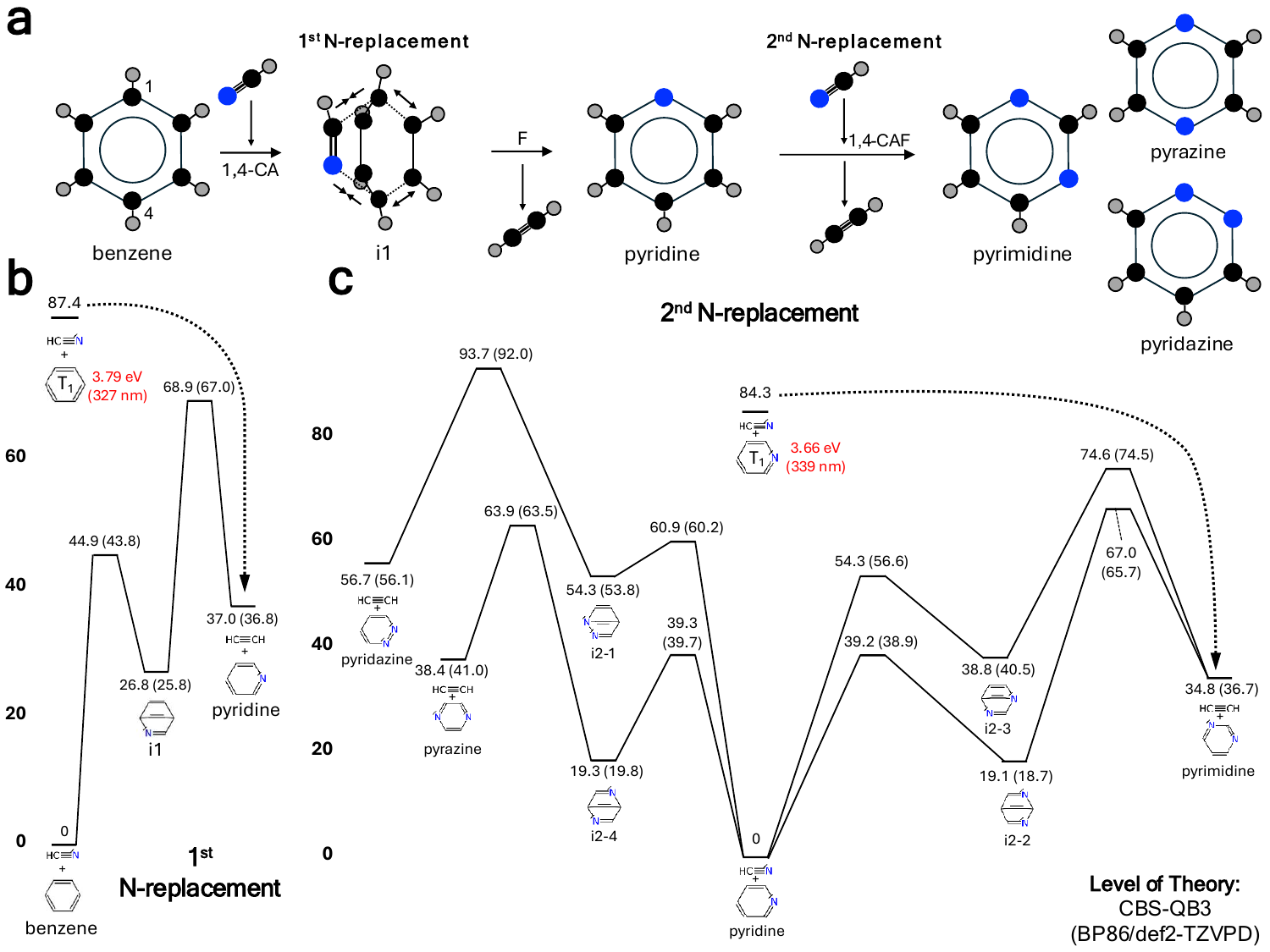}
\caption{\textbf{Overall information on the first and second nitrogen replacements in a benzene ring.} \textbf{a}. The schematic diagram illustrating the formation of \ce{C4H4N2} isomers (pyrimidine, pyrazine, and pyridazine) via 1,4-cycloaddition and fragmentation (CAF) from the benzene + HCN reaction. Each colored ball represents a different atomic species: black for carbon, gray for hydrogen, and blue for nitrogen. CA stands for cycloaddition, and F stands for fragmentation. \textbf{b}. The \ce{C7H7N} potential energy surface (first nitrogen replacement). \textbf{c}. The \ce{C6H6N2} potential energy surface (second nitrogen replacement). Numbers indicate potential energy (units in kcal/mol) calculated at the CBS-QB3 level of theory, while numbers in parentheses indicate potential energy calculated at the BP86/def2-TZVPD level of theory, considering water solvation effect at 298 K. The red numbers indicate the equivalent energy (in eV) and the corresponding wavelength (in nm) of the energy values shown in kcal/mol (black numbers) for each photoexcited triplet state (T$_1$) of benzene and pyridine. The black dotted lines indicate well-skipping reactions which lead to the formation of pyridine+\ce{C2H2} and pyrimidine+\ce{C2H2}, respectively.}\label{fig: fig_pyrimidine}
\end{figure*}

Both \ce{C6H6} and \ce{HCN} are abundant in various planetary atmospheres (e.g., Titan \citep{COUSTENIS2003383}), and both possess $\pi$-bonds. Unlike classical 1,4-cycloaddition reactions involving highly strained dienophiles such as o-benzyne, the present system involves hydrogen cyanide as the reacting species. Although HCN does not possess comparable ring strain, our \textit{ab initio} calculations using the methodology detailed in Section~\ref{sec: gaussian 09} show that the reaction becomes accessible as shown in Figure~\ref{fig: fig_pyrimidine}. The feasibility of this mechanism is supported by explicit transition state searches and calculated activation barriers, which are provided as \texttt{Gaussian 09} output files in supplementary information. A schematic illustrating nitrogen incorporation into the aromatic ring via the 1,4-CAF mechanism is shown in Figure~~\ref{fig: fig_pyrimidine}a. As shown in Figure~ \ref{fig: fig_pyrimidine}b, the first nitrogen substitution occurs via the formation of the intermediate species i1 from benzene + HCN reaction, with an initial reaction barrier of 44.9 kcal/mol, followed by a second barrier of 68.9 kcal/mol leading to pyridine (\ce{C5H5N}) and \ce{C2H2}. Due to the symmetric geometry of the benzene ring, pyridine is the only possible product regardless of orientation. The kinetic information for these reactions (i.e., the corresponding rate coefficients) is provided in parentheses in Table~\ref{tab: rate_coefficients} for use in kinetic modeling studies, and is also included in the Supplementary Information in \texttt{CHEMKIN} format.

Pyridine can subsequently react with additional HCN, forming four different isomers (i.e., i2-1, i2-2, i2-3, and i2-4), which ultimately lead to three \ce{C4H4N2} isomers: pyrimidine, pyrazine, and pyridazine, as shown in Figure~ \ref{fig: fig_pyrimidine}c. Unlike the first nitrogen substitution, the orientation of subsequent additions matters. Based on computed reaction barriers, pyrazine formation through i2-4 (highest barrier: 63.9 kcal/mol) and pyrimidine formation through i2-2 (highest barrier: 67.0 kcal/mol) are the most favorable pathways. Consequently, the minimum energy required to overcome the highest reaction barrier in the full sequence from benzene to pyrimidine is 68.9 kcal/mol, corresponding to a photon energy of 415 nm, near the boundary between the UV-A and visible wavelength range. The kinetic information for these reactions (i.e., the corresponding rate coefficients) is provided in parentheses in Table~\ref{tab: rate_coefficients} for use in kinetic modeling studies, and is also included in the Supplementary Information in \texttt{CHEMKIN} format.

Notably, this reaction barrier is significantly lower than that required for HCN oligomerization, as the monomer-to-dimer step requires 80 kcal/mol \citep{benallou2016understanding}, and the dimer-to-trimer step requires 65 kcal/mol \citep{benallou2017mechanism}. However, the reaction barrier is still high to efficiently proceed under purely thermal conditions due to the aromatic stability of benzene. If one considers that this reaction occurs on the surface of a liquid water ocean, the ``on water'' catalysis effect \citep{narayan2005water} (or the solvent effect) might play a crucial role; the reaction may also proceed under aqueous conditions in bulk water, where solvent effects could similarly influence the reaction. This solvent effect, occurring at the oil-water interface, has been shown to enhance Diels-Alder-type reaction rates by up to five orders of magnitude at 300 K \citep{narayan2005water}. This enhancement is attributed to OH groups forming stronger hydrogen bonds with the transition state than with the reactants \citep{jung2007theory} and could similarly enhance the rate of our proposed reaction pathway. To evaluate the extent to which solvation can lower the activation energy, we calculated kinetic solvent effects using a hybrid quantum chemical and \texttt{COSMO-RS} approach \citep{chung2023solvent} (see method details in Section \ref{sec: cosmors}). As shown in Figure~ \ref{fig: fig_pyrimidine}b and c, values in parentheses represent the potential energy surface with solvation effects included (using water as the solvent at 298 K). The solvent effect reduces the activation energy by only about 1 kcal/mol, which suggests that additional catalytic or environmental mechanisms are likely required to make this reaction feasible under the relatively low-temperature conditions of early Earth or early Mars.

One promising driving force is the involvement of metastable benzene and pyridine generated via UV photoexcitation, followed by intersystem crossing. Upon UV absorption, benzene and pyridine are excited to their first singlet excited states and can undergo intersystem crossing to isoenergetic triplet states \citep{duncan1981photoionization, knight1976radiative}. Their collision-free lifetimes (470 ns for benzene \citep{duncan1981photoionization} and 1$\mu$s for pyridine \citep{terazima1988quantum}) are sufficiently long to allow collisions with HCN, providing enough internal energy to overcome the reaction barriers discussed above and ultimately leading to the formation of pyridine and pyrimidine (i.e., well-skipping reaction indicated in Figure~ \ref{fig: fig_pyrimidine}b and c). We evaluated the potential of this photoexcitation pathway to form heterogeneous aromatic species on early Mars by employing photochemical kinetic-transport modeling combined with the photoexcitation framework described in \citet{yang2023high}. Key rate coefficients are summarized in Table~\ref{tab: rate_coefficients}, with detailed methods provided in Sections~\ref{sec: KINETICS} and \ref{sec: arkane}. Results are presented and discussed in Section~\ref{sec:mars}.

Overall, the 1,4-CAF reaction provides an efficient pathway for pyrimidine formation, offering a plausible upstream route to the heterocyclic building block required for pyrimidine-type nucleobases (i.e., thymine, cytosine, and uracil). Further experimental studies are strongly recommended to evaluate the reaction rates of the 1,4-CAF process between benzene and HCN and its role in pyridine and pyrimidine formation. For example, high-temperature shock tube experiments could probe the kinetics of thermally activated reactions ($\sim$1500 K), while photochemical experiments under aqueous conditions, such as ultraviolet irradiation of benzene in the presence of HCN dissolved in water, could test whether photoexcitation and solvent effects facilitate this pathway in environments relevant to prebiotic chemistry.

\begin{table*}[htb!]
\centering
\small
\begin{threeparttable}

\caption{\textbf{Key rate coefficients involving photoexcited metastable species (i.e., $^3$\ce{C6H6} or $^3$\ce{C5H5N})} used to simulate pyrimidine formation at the surface of the early Martian atmosphere (Section~\ref{sec:mars}). Note that the numbers in parentheses represent the rate coefficients for reactions involving ground-state benzene or pyridine (i.e., $^1$\ce{C6H6} or $^1$\ce{C5H5N}), calculated based on the potential energy surfaces described in Section~\ref{sec: gaussian 09} and shown in Figure~\ref{fig: fig_pyrimidine}. These rate coefficients are $T,P-$dependent, following a pressure-dependent Arrhenius form detailed in Section 2.3.2(3) of \citet{yang2024automated}. Details on importing rate coefficients from potential energy surface calculations are in Section~\ref{sec: arkane}; remaining rate coefficients for chemical kinetic modeling are provided in the supplementary information.}
\label{tab: rate_coefficients}

\begin{tabular}{@{}llll@{}}
\toprule
\multicolumn{1}{c}{$P$ [atm]} &
\multicolumn{1}{c}{$A$\tnote{a}} &
\multicolumn{1}{c}{$n$} &
\multicolumn{1}{c}{$E_a$ [kcal/mol]} \\
\midrule

\multicolumn{4}{c}{\textbf{$^1$\ce{C6H6} $\rightarrow$ $^3$\ce{C6H6} ($\Phi_{\rm ISC}$=0.8)}\tnote{b}} \\
[4pt]
\multicolumn{1}{c}{--} & \multicolumn{1}{c}{1.170$\times10^{-5}$} & \multicolumn{1}{c}{--} & \multicolumn{1}{c}{--} \\
\midrule

\multicolumn{4}{c}{\textbf{$^3$\ce{C6H6} $\rightarrow$ $^1$\ce{C6H6} ($\tau$=470 ns)}\tnote{c}} \\
[4pt]
\multicolumn{1}{c}{--} & \multicolumn{1}{c}{2.128$\times10^{6}$} & \multicolumn{1}{c}{--} & \multicolumn{1}{c}{--} \\
\midrule

\multicolumn{4}{c}{\textbf{$^3$\ce{C6H6} + HCN $\rightarrow$ \ce{C5H5N} + \ce{C2H2}}} \\
[4pt]
0.001316   & 2.757$\times10^{2}$ (1.033$\times10^{3}$)     & 1.719 (2.673)  & -1.768 (67.248)  \\
0.028657   & 2.502$\times10^{5}$ (1.033$\times10^{3}$)     & 0.910 (2.673)  & 0.490 (67.248)   \\
0.624134   & 1.150$\times10^{7}$ (1.033$\times10^{3}$)     & 0.506 (2.673)  & 3.297 (67.248)   \\
13.593250  & 9.989$\times10^{2}$ (2.140$\times10^{3}$)     & 1.729 (2.586)  & 3.906 (67.462)   \\
296.052632 & 1.979$\times10^{-6}$ (3.607$\times10^{5}$)    & 4.229 (1.997)  & 2.344 (69.764)   \\
\midrule

\multicolumn{4}{c}{\textbf{$^1$\ce{C5H5N} $\rightarrow$ $^3$\ce{C5H5N} ($\Phi_{\rm ISC}$=0.5)}\tnote{d}} \\
[4pt]
\multicolumn{1}{c}{--} & \multicolumn{1}{c}{2.616$\times10^{-4}$} & \multicolumn{1}{c}{--} & \multicolumn{1}{c}{--} \\
\midrule

\multicolumn{4}{c}{\textbf{$^3$\ce{C5H5N} $\rightarrow$ $^1$\ce{C5H5N} ($\tau$=1 $\mu$s)}\tnote{e}} \\
[4pt]
\multicolumn{1}{c}{--} & \multicolumn{1}{c}{1.000$\times10^{6}$} & \multicolumn{1}{c}{--} & \multicolumn{1}{c}{--} \\
\midrule

\multicolumn{4}{c}{\textbf{$^3$\ce{C5H5N} + HCN $\rightarrow$ pyridazine + \ce{C2H2}}} \\
[4pt]
0.001316   & 1.351$\times10^{9}$ (4.241$\times10^{1}$)     & 0.001 (2.888)  & 9.832 (91.941)   \\
0.028657   & 5.828$\times10^{12}$ (4.241$\times10^{1}$)    & -0.988 (2.888) & 12.794 (91.941)  \\
0.624134   & 3.855$\times10^{13}$ (4.241$\times10^{1}$)    & -1.149 (2.888) & 15.347 (91.941)  \\
13.593250  & 4.760$\times10^{8}$ (4.240$\times10^{1}$)     & 0.312 (2.888)  & 15.588 (91.941)  \\
296.052632 & 3.653$\times10^{-1}$ (4.240$\times10^{1}$)    & 2.923 (2.888)  & 13.818 (91.941)  \\
\midrule

\multicolumn{4}{c}{\textbf{$^3$\ce{C5H5N} + HCN $\rightarrow$ pyrimidine + \ce{C2H2}}} \\
[4pt]
0.001316   & 2.683$\times10^{4}$ (8.183$\times10^{-3}$)    & 1.279 (3.716)  & -0.528 (63.677)  \\
0.028657   & 4.284$\times10^{6}$ (8.183$\times10^{-3}$)    & 0.696 (3.716)  & 1.769 (63.677)   \\
0.624134   & 2.173$\times10^{7}$ (8.183$\times10^{-3}$)    & 0.558 (3.716)  & 4.148 (63.677)   \\
13.593250  & 2.933$\times10^{3}$ (8.838$\times10^{-3}$)    & 1.715 (3.706)  & 4.775 (63.700)   \\
296.052632 & 1.835$\times10^{-5}$ (5.005$\times10^{-1}$)   & 4.075 (3.236)  & 3.440 (65.299)   \\
\midrule

\multicolumn{4}{c}{\textbf{$^3$\ce{C5H5N} + HCN $\rightarrow$ pyrazine + \ce{C2H2}}} \\
[4pt]
0.001316   & 4.584$\times10^{2}$ (4.175$\times10^{1}$)     & 1.680 (2.783)  & -1.737 (62.164)  \\
0.028657   & 2.357$\times10^{5}$ (4.175$\times10^{1}$)     & 0.940 (2.783)  & 0.419 (62.164)   \\
0.624134   & 2.360$\times10^{6}$ (4.175$\times10^{1}$)     & 0.726 (2.783)  & 2.961 (62.164)   \\
13.593250  & 2.853$\times10^{1}$ (5.693$\times10^{1}$)     & 2.186 (2.745)  & 3.180 (62.253)   \\
296.052632 & 1.471$\times10^{-8}$ (8.076$\times10^{3}$)    & 4.839 (2.169)  & 1.303 (64.270)   \\

\bottomrule
\end{tabular}

\begin{tablenotes}[flushleft]
\footnotesize
\item[a] [s$^{-1}$] for a unimolecular reaction; [cm$^{3}$/mol/s] for a bimolecular reaction.
\item[b] The intersystem crossing quantum yield ($\Phi_{\rm ISC}$) taken from \citet{duncan1981photoionization}.
\item[c] Collision-free lifetime ($\tau$) of 470 ns was taken from \citet{duncan1981photoionization}.
\item[d] $\Phi_{\rm ISC}$=0.5 was taken from \citet{knight1976radiative}.
\item[e] $\tau$=1 $\mu$s was taken from \citet{terazima1988quantum}.
\end{tablenotes}

\end{threeparttable}
\end{table*}

\subsection{Formation of Purine (\ce{C5H4N4}) from Pyrimidine (\ce{C4H4N2})}\label{sec:purine}
\begin{figure*}[hbt!]
\centering
\includegraphics[width=1\textwidth]{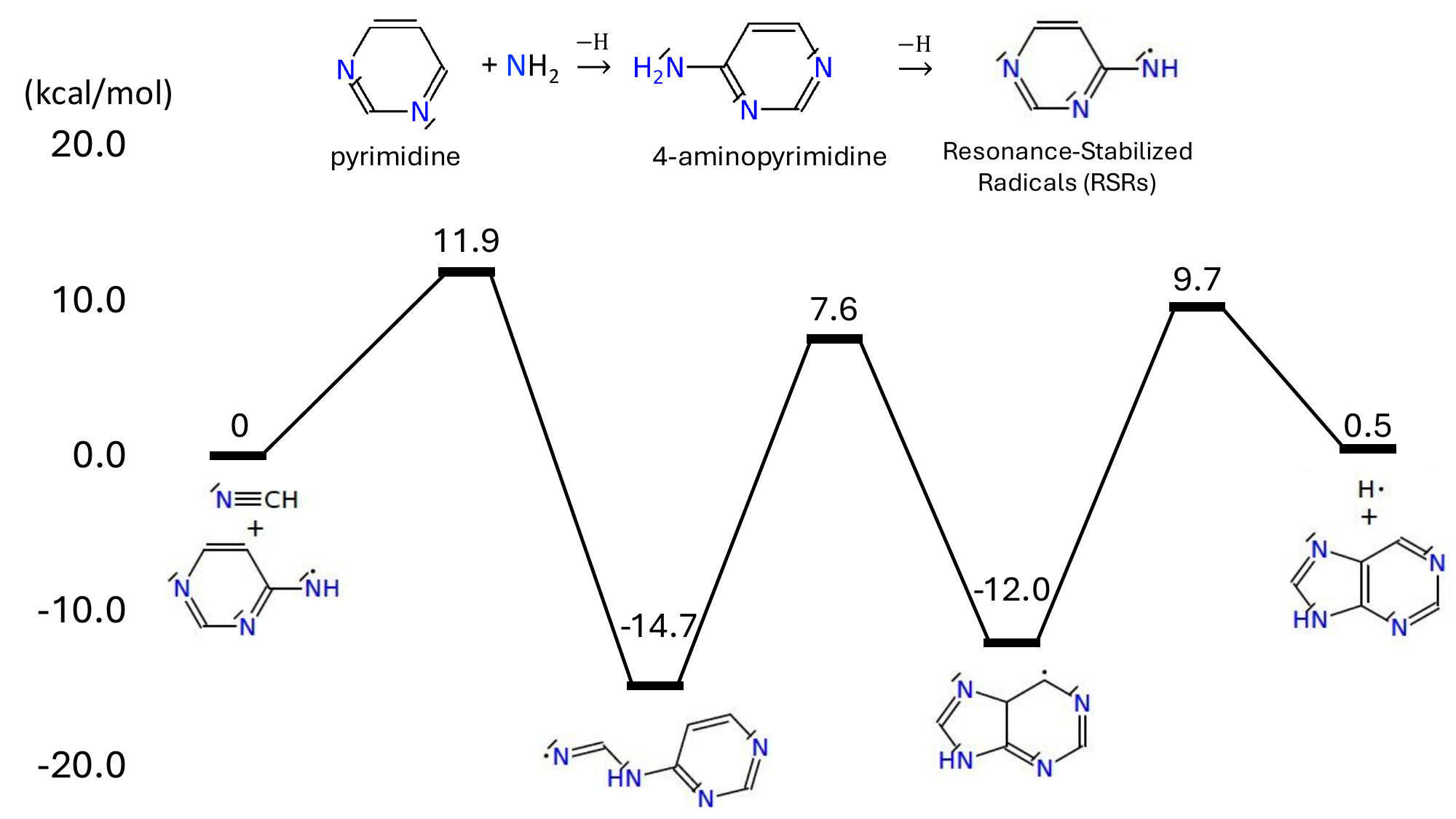}
\caption{\textbf{Overall information on the purine formation from the pyrimidine + \ce{NH3} + HCN reaction.} The \ce{C5H5N4} potential energy surface is calculated at the CBS-QB3 level of theory, with units in kcal/mol.}\label{fig: fig_purine}
\end{figure*}

Pyrimidine can further lead to the formation of purine, as shown in Figure~ \ref{fig: fig_purine}. The first step involves the reaction of pyrimidine with an amino radical (\ce{NH2}), which may arise from photodissociation of \ce{NH3} or from aqueous nitrogen chemistry, resulting in the formation of 4-aminopyrimidine. A theoretical study on the reaction between benzene and \ce{NH2} \citep{altarawneh2021updated} reports that the addition pathway proceeds with a reaction barrier of 11.23 kcal/mol, ultimately yielding aniline (\ce{C6H5NH2}). Given the structural similarity between pyrimidine+\ce{NH2} and benzene+\ce{NH2}, the reaction mechanism is expected to be analogous, leading to the formation of 4-aminopyrimidine, as illustrated in Figure~ \ref{fig: fig_purine}. While amino radicals could react with multiple molecular species present in the environment, quantitatively assessing the competition among these reaction pathways remains an important topic for future modeling and is beyond the scope of the present study.

If 4-aminopyrimidine subsequently undergoes hydrogen loss at the amine site, it forms a resonance-stabilized radical (RSR). The subsequent addition of hydrogen cyanide (\ce{HCN}) to this radical has been investigated using \textit{ab initio} calculations (Figure~ \ref{fig: fig_purine}). This reaction proceeds via an entrance barrier of 11.9 kcal/mol, followed by two intermediates with barriers of 7.6 kcal/mol and 9.7 kcal/mol, before ultimately yielding purine and an H radical as final products. Purine may, in principle, undergo further functionalization by species such as (\ce{OH}-), amine (\ce{NH3}-), and H-containing reactants to yield purine-type nucleobases (i.e., adenine and guanine), although the position-specific pathways and selectivity of these downstream reactions are beyond the scope of the present study.

It should be emphasized that the present work establishes pathways to the heterocyclic cores pyrimidine and purine, but does not explicitly resolve the downstream, position-specific functionalization steps required to form the canonical nucleobases (i.e., adenine, guanine, cytosine, thymine, and uracil). Because such downstream reactions may yield multiple constitutional isomers depending on regioselectivity and competing pathways, their detailed formation pathways (if not one) remain an important direction for future work.

Considering the solubility of pyrimidine, ammonia, and 4-aminopyrimidine in water, aqueous environments could facilitate the formation of 4-aminopyrimidine from \ce{NH3} and pyrimidine, as well as the subsequent HCN addition leading to purine, by allowing HCN and other reactants to dissolve and accumulate in the liquid phase, thereby increasing their effective concentrations, while solvent effects may further stabilize intermediates or transition states. This theoretical study strongly highlights the potential role of air–water interfacial chemistry between benzene and cyanide in the formation of nucleobase precursors, pyrimidine, and purine. These findings strongly support the need for immediate follow-up studies to further investigate these reaction pathways under early Earth-like conditions.

\subsection{1D Photochemical Modeling of Early Martian Atmosphere}\label{sec:early_martian}

\begin{figure*}[hbt!]
\centering
\includegraphics[width=1\textwidth]{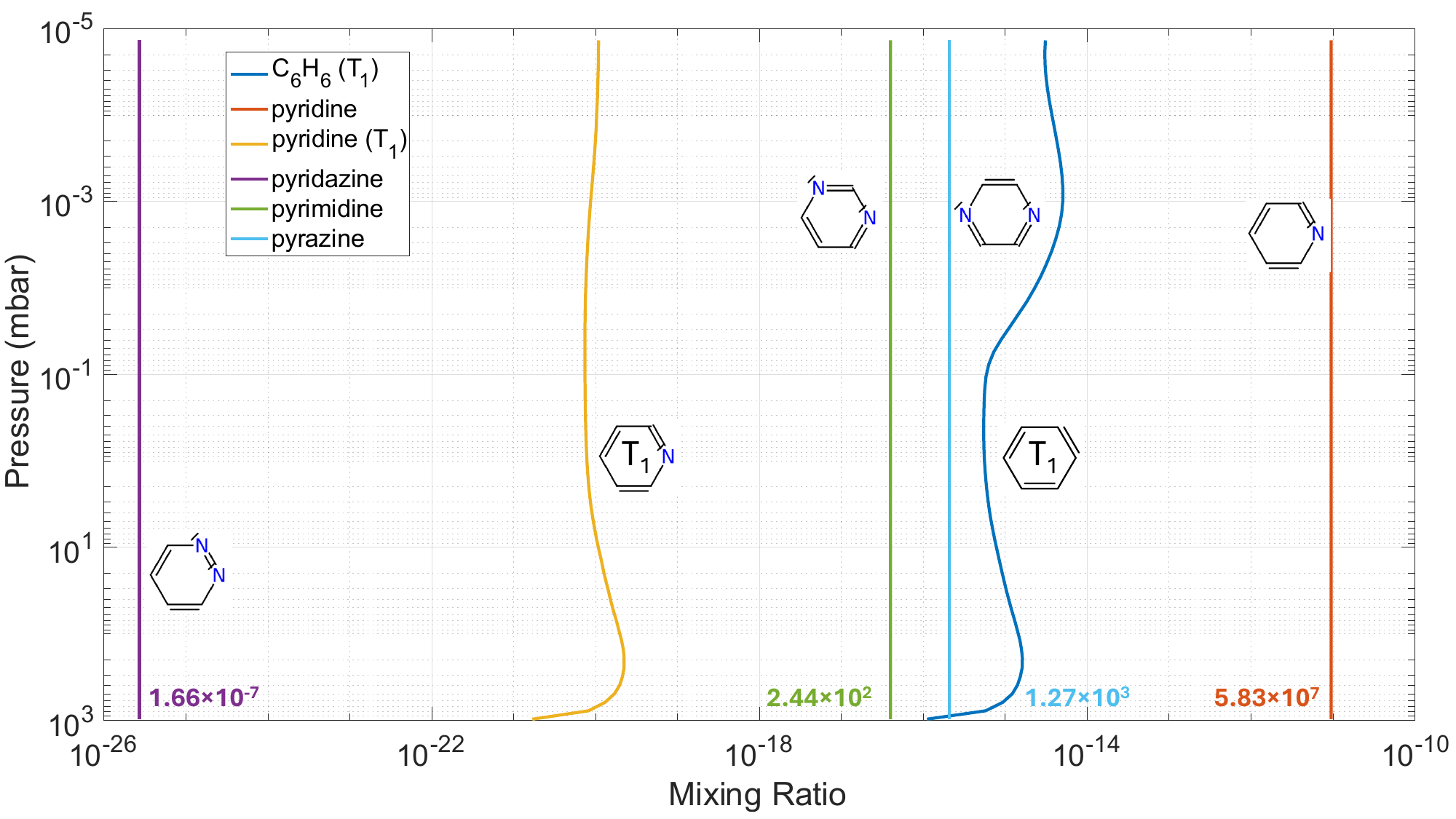}
\caption{\textbf{Vertical mixing ratio profiles of heterocyclic aromatic species from 1D Photochemical Modeling of the Early Martian Atmosphere During the Cold and Dry Period described in \citet{adams2025episodic}.} Each colored solid line represents a different species. The corresponding colored numbers indicate surface deposition rates calculated by \texttt{KINETICS}, with units of [molecules/cm$^{2}$/s]}\label{fig: Martiansurface}
\end{figure*}

Figure~~\ref{fig: Martiansurface} presents the results of the 1D photochemical kinetic–transport atmospheric modeling of the early Martian atmosphere described in Section~\ref{sec: KINETICS}. The predicted pyridine mixing ratio is approximately $10^{-11}$, followed by photoexcited triplet-state benzene ($^3$\ce{C6H6}), which decreases from $\sim10^{-14}$ in the upper atmosphere to $\sim10^{-16}$ near the surface. This decreasing mixing ratio with decreasing altitude is due to the increasing UV opacity of the atmosphere, as illustrated in Figure~~\ref{photons}. Among the \ce{C4H4N2} isomers, pyrazine is predicted to be the most abundant ($\sim10^{-16}$), followed by pyrimidine at roughly an order of magnitude lower abundance ($\sim10^{-17}$). The mixing ratio of pyridazine is predicted to be significantly smaller (approximately $10^{-26}$), even lower than the photoexcited triplet-state pyridine ($\sim10^{-20}$). This very low abundance of pyridazine is consistent with its high reaction barrier (93.7 kcal/mol), even when assuming a reaction of photoexcited pyridine with HCN, as depicted in Figure~\ref{fig: fig_pyrimidine}c. If this photoexcitation mechanism indeed represents the dominant prebiotic pathway for forming heterocyclic aromatic species from benzene and HCN on the early Martian surface, pyrazine derivatives are expected to be present at concentrations roughly an order of magnitude higher than pyrimidine derivatives, while pyridazine derivatives would be much less detected compared to the other two isomers.

Assuming a dry deposition velocity ($V_{\rm dep}$) of 0.2 cm/s, based on an analogy to benzene \citep{cohen-1994} due to the aromatic character of the molecules considered here, surface deposition rates [molecules/cm$^{2}$/s] were computed using \texttt{KINETICS}, as shown in Figure~~\ref{fig: Martiansurface}. Using a mean Martian radius of 3389.5 km \citep{archinal2018report_mars_radius}, we estimate the following annual mass deposition rates: 2.77$\times10^{7}$ kg/yr for pyridine (\ce{C5H5N}); 1.18$\times10^{2}$ kg/yr for pyrimidine; 6.10$\times10^{2}$ kg/yr for pyrazine; and 7.98$\times10^{-3}$ kg/yr for pyridazine. These numbers are comparable to previous theoretical estimates of the required delivery flux of intact carbon to support prebiotic chemistry in the early Earth, primarily through interplanetary dust particles (IDPs) and meteorites, estimated at approximately $\sim6\times10^7$ kg/yr and $\sim2\times10^3$ kg/yr, respectively \citep{chyba1992endogenous, pearce2017origin}. The modeled deposition rates from our photoexcitation-driven pathway, particularly for pyridine, fall within or exceed these estimates, suggesting that atmospheric in situ production could have served as a significant endogenous source of nucleobases and nucleobase precursors.  If pyridine \citep[completely miscible in water;][]{haynes2016crc} were to dissolve into liquid water, become locally concentrated through processes such as wet–dry cycles, and subsequently react with similarly concentrated HCN in the presence of an energy source such as UV radiation or lightning, this could lead to the further efficient formation of pyrimidine and other nitrogen-substituted aromatic species (e.g., purine). These results suggest that photochemically driven local processes may represent one possible pathway contributing to prebiotic chemistry on early Mars (or Earth), enabled by the reactions we propose starting from benzene and HCN via the 1,4-CAF mechanism.

It should be noted that several parameters in the photochemical modeling, such as dry deposition velocities and the lower boundary methane mixing ratio, are not well constrained for early Martian environments and therefore require simplifying assumptions. A comprehensive sensitivity analysis exploring the full parameter space would be valuable but lies beyond the scope of the present study.

\section{Discussions}\label{discussion}
\subsection{Implications for Hadean Earth Atmosphere}\label{sec:hadean}
\begin{figure*}[hbt!]
\centering
\includegraphics[width=1\textwidth]{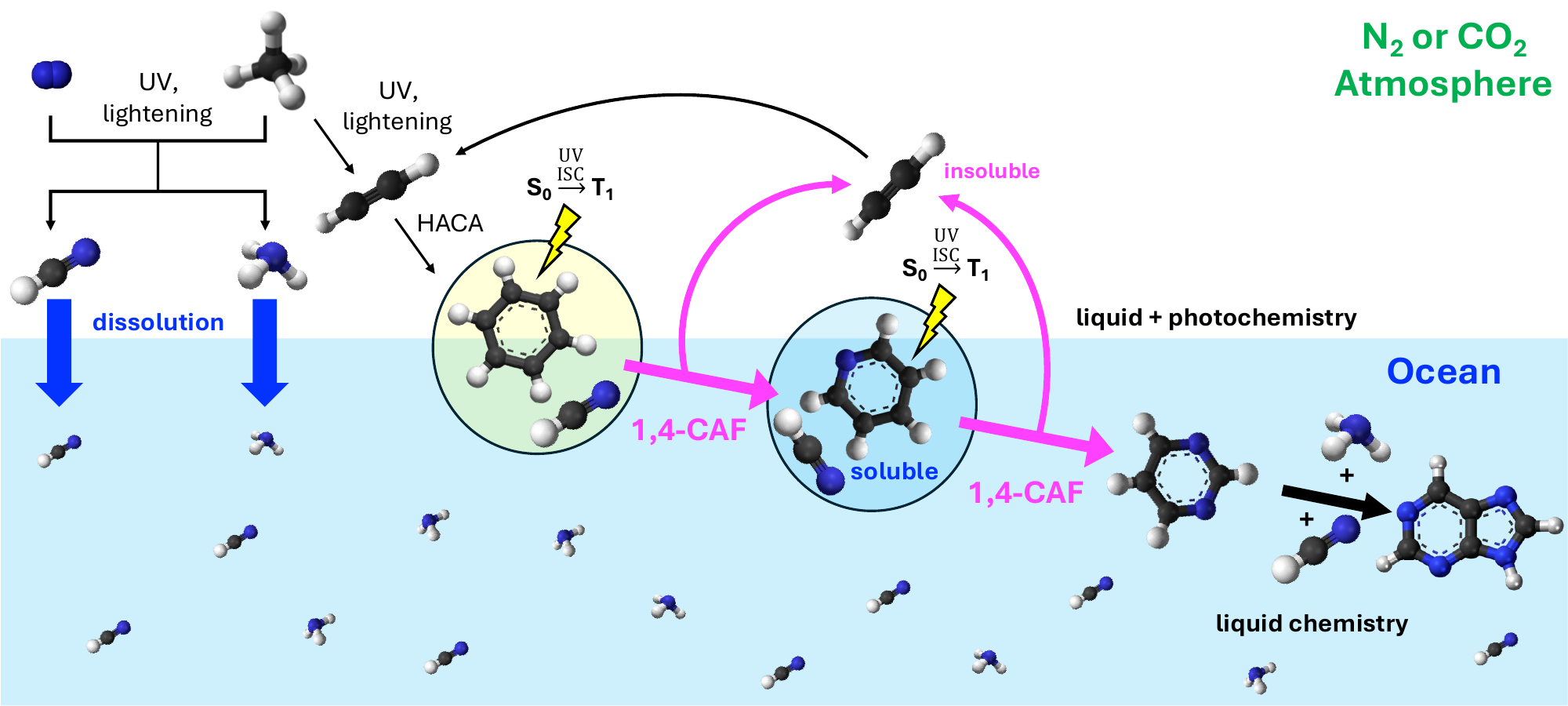}
\caption{\textbf{Schematic illustration of the Hadean Earth prebiotic chemistry cycle.} Small molecules, ranging from HCN and \ce{NH3} to benzene, are synthesized via atmospheric photochemistry and lightning, while heterocyclic aromatic species form on the ocean surface aided by UV-photoexcitation and 1,4-CAF reaction, ultimately leading to pyrimidine and purine synthesis. CAF refers to cycloaddition and fragmentation.}\label{fig: HadeanEarthAtmosphere}
\end{figure*}

We propose a prebiotic nucleobase formation in the Hadean Earth atmosphere based on our newly proposed pyrimidine and purine synthesis from the benzene + HCN reaction. As shown in Figure~ \ref{fig: HadeanEarthAtmosphere}, high-energy sources such as far-ultraviolet (FUV) photochemistry and ion-driven chemistry (e.g., lightning) sequentially produce HCN \citep{dodonova1966activation, scattergood1989production}, \ce{NH3} \citep{bernard2003experimental, yelle2010formation}, and \ce{C2H2} \citep{romani1993methane}. These species are abundant and well detected in Titan's \ce{N2}-dominated atmosphere \citep{COUSTENIS2003383, teanby2007vertical, vuitton2006nitrogen}, which contains 5\% \ce{CH4}, a composition also believed to be available in early Earth \citep{urey1952early, zahnle2020creation}.

However, unlike Titan, early Earth had a liquid phase water ocean \citep{wilde2001evidence, sleep2001initiation, valley2002cool}, where HCN and \ce{NH3} readily dissolved, with Henry's law constants of $\sim$8.9$\times10^{-2}$ [mol$\cdot$m$^{-3}\cdot$Pa$^{-1}$] for HCN and 5.9$\times10^{-1}$ [mol$\cdot$m$^{-3}\cdot$Pa$^{-1}$] for \ce{NH3} \citep{burkholder2020chemical}, although their solubility depends on pH (pKa$\sim9.21$ for HCN and pKa$\sim9.25$ for \ce{NH3} \citep{haynes2016crc}, while \ce{C2H2} remained in the atmosphere. This \ce{C2H2} might have continuously formed benzene (\ce{C6H6}) via hydrogen-abstraction acetylene-addition (HACA) chemistry \citep{bittner1981composition, frenklach1985detailed} driven by UV photons. Given benzene's thermal stability under early Earth conditions, and its low solubility in water, \ce{C6H6} would likely remain primarily in the atmosphere, although interactions with the ocean surface may also occur. The 1,4-cycloaddition and fragmentation (CAF) reaction between atmospheric \ce{C6H6} and ocean-dissolved HCN might have been accelerated by the combination of ``on water'' catalysis effect \citep{narayan2005water} and UV-photoexcitation-aided reactions mentioned in Section \ref{sec:pyridine}, and could lead to the first nitrogen substitution in the benzene ring, forming pyridine, along with \ce{C2H2}.

Due to its dipole moment from nitrogen substitution, pyridine is water-miscible, while \ce{C2H2} is released back into the atmosphere, where it recycles into benzene via atmospheric photochemistry. Subsequently, pyridine would continuously react with HCN in liquid-phase chemistry, augmented by UV-photoexcitation, forming pyrimidine, the fundamental building blocks of pyrimidine-type nucleobases (i.e., cytosine, thymine, and uracil). At sufficiently high concentrations, pyridine may also absorb UV radiation and attenuate photon penetration, potentially limiting photoexcitation to shallower layers. Quantifying this self-shielding effect is beyond the scope of this work. Pyrimidine would then react with \ce{NH3} to form 4-aminopyrimidine, which subsequently reacts with HCN to form purine, the precursor for purine-type nucleobases (i.e., adenine and guanine). However, the accumulation of prebiotic molecules in natural environments faces well-known challenges, including dilution, competing side reactions, and degradation processes that can limit the buildup of complex organics \citep{walton2022can, white2024nothing}. In this work, our goal is not to predict absolute concentrations of these intermediates, but rather to identify quantum-chemically viable reaction pathways and chemically stable intermediates that could participate in nucleobase formation under plausible early Earth conditions.

A detailed quantitative treatment of these processes is beyond the scope of the present study. Modeling prebiotic chemistry on early Earth is substantially more complex than modeling purely gas-phase atmospheric chemistry, because condensed-phase reactions within the ocean and ocean–atmosphere interactions likely play important roles. Many of these catalytic and condensed-phase processes remain poorly constrained, which currently limits precise kinetic modeling. Future studies aimed at determining the rates of these reactions will therefore be important for assessing the potential accumulation of nucleobases in early Earth environments.

\subsection{Implications for early Martian Atmosphere}\label{sec:mars}
Our model simulates heterocyclic aromatic compound formation under cold, dry surface conditions on early Mars, conditions that favor high benzene and HCN concentrations but lack liquid water. This scenario corresponds to the cold, CO-rich epochs described by Adams \textit{et al.,} 2025, which alternate with transient warm and wet periods driven by episodic \ce{H2} outgassing. While \citet{adams2025episodic} does not explicitly address the exact duration of individual cold, dry epochs and their corresponding behavior of organic molecules, we propose that heterocyclic organics formed during dry periods and deposited on the Martian surface may have subsequently condensed or dissolved into surface waters during warmer episodes. However, the efficiency of this process would depend on compound-specific solubility, as heterocyclic organics span a wide range of aqueous solubilities. For example, some nucleobases \citep[e.g., guanine;][]{devoe1984aqueous} are relatively insoluble under neutral conditions, whereas others are more readily dissolved, suggesting selective partitioning into aqueous environments. Such transitions could facilitate the accumulation and preservation of these compounds in sedimentary deposits. This suggests that Mars Sample Return efforts should prioritize locations with geologic evidence of past liquid water, particularly sedimentary environments where photochemically produced organics may have been buried and sequestered, thereby reducing exposure to surface-driven degradation processes such as UV radiation and oxidation.

These findings offer a valuable framework for interpreting potential organic signatures in returned Martian samples, particularly possible prebiosignatures \citep{summons2011preservation, eigenbrode2018organic}, defined as molecular indicators of abiotic chemical processes that may precede the emergence of life. If heterocyclic aromatic compounds such as pyrimidine or pyrazine are detected in ancient sediments, their presence could support the hypothesis that photoexcitation-driven surface chemistry played a role in prebiotic molecule formation on early Mars. Similar classes of molecules have been proposed as potential prebiosignatures in planetary environments, reflecting chemical pathways that may operate prior to the emergence of biological activity \citep[e.g.,][]{cleaves2008reassessment, sutherland2017opinion}. The predicted deposition rates outlined here provide testable constraints that can guide future sample selection, extraction strategies, and in situ analyses. As the Mars Sample Return mission advances toward its 2030 launch, modeling efforts like this can help prioritize landing sites with a greater likelihood of preserving photochemically produced organics.

\subsection{Impact Events as a Potential Driving Force}\label{sec:impact}
Many previous studies suggest that extraterrestrial objects impacted the early Earth intensively \citep{culler2000lunar, hartmann2000time, valley2002cool}, delivering not only large amounts of material but also substantial amounts of energy to the surface environment. Although this study has focused on ultraviolet photon energy from the Sun as a driving force for the proposed chemical mechanisms in the early Earth, the episodic and intense energy released during large impact events represents an additional high-energy source that could have driven prebiotic chemistry. Numerous laboratory studies motivated by this idea have subjected gas mixtures or solid materials to shock heating followed by rapid thermal quenching, resulting in the formation of products ranging from simple amino acids and nucleobases \citep{bar1970shock, chyba1992endogenous, furukawa2015nucleobase, ferus2020one} to complex organic agglomerates \citep{singh2020shock, surendra2021complex}. In such experiments, the incident shock front can raise temperatures up to 4000 K \citep{bar1970shock}, which is more than sufficient to overcome the highest reaction barrier of 68.9 kcal/mol required for pyrimidine formation from benzene along the potential energy surfaces shown in Figure~\ref{fig: fig_pyrimidine}. However, it has to be noted that impact events may also lead to the thermal decomposition of complex organic molecules. Experimental studies have shown that benzene can be efficiently destroyed during impacts in \ce{N2}-dominated atmospheres \citep{petera2023decomposition}. Thus, impact-driven environments may both generate and destroy aromatic precursors, with the net outcome depending on local temperature, atmospheric composition, and cooling timescales.

When compared with the photoexcited metastable augmented rate coefficients derived in Table~\ref{tab: rate_coefficients}, for example for the \ce{C6H6} + HCN $\rightarrow$ \ce{C5H5N} + \ce{C2H2} reaction at 0.62 atm, the impact-driven reaction rate becomes comparable at approximately 3500 K ($k \sim 4.5 \times 10^{8}$ cm$^{3}$ mol$^{-1}$ s$^{-1}$) and nearly twice as fast as the photoexcitation-driven rate at 4000 K. This suggests that sufficiently large impacts capable of raising local temperatures above several thousand Kelvin could efficiently drive the proposed nitrogen incorporation pathways. In such a scenario, continuous photochemistry may act to accumulate benzene and HCN in surface reservoirs, while episodic impacts provide brief but intense thermal conditions that promote rapid formation of nitrogen-substituted aromatic species. Future work should quantitatively link impactor size, thermal histories, and chemical yields to assess the relative importance of this mechanism.

\section{Conclusion}\label{sec:conclusion}

In this work, we combined state-of-the-art computational chemistry methods, including automated chemical network generation, \textit{ab initio} quantum chemistry, and one-dimensional photochemical–transport modeling, to identify plausible chemical pathways for the formation of core nucleobase precursors in DNA and RNA. We find that benzene can be a stable and abundant aromatic reservoir in \ce{N2}-dominated or \ce{CO2}-dominated atmospheres and can serve as an upstream building block for prebiotic nitrogen incorporation when combined with HCN through a simple and efficient heteroaromatic formation mechanism involving HCN addition to aromatic $\pi$ systems. Quantum chemistry calculations show that benzene reacts with HCN through a 1,4-HCN-cycloaddition/\ce{C2H2}-fragmentation (CAF) reaction to yield pyridine and \ce{C2H2}, and that subsequent HCN addition to pyridine produces \ce{C4H4N2} isomers, including pyrimidine. These reaction pathways can be activated by either ultraviolet photoexcitation or episodic impact-driven thermal energy, and we extend this framework to implications for both early Earth and early Mars. In particular, our model suggests that the proposed reaction pathways may have served as a significant endogenous source of nucleobase precursors, with implications for prebiotic chemistry. Overall, the benzene + HCN CAF mechanism provides a mechanistically explicit route to pyrimidine and purine precursors under anoxic conditions, linking atmospheric chemistry to surface reservoirs on early Earth and early Mars and offering quantitative targets for future laboratory validation and sample-return interpretation.

\section*{Acknowledgements}
The authors gratefully acknowledge Dr. Jonathan Zheng and Professor William H. Green Jr. at the Massachusetts Institute of Technology for training Jeehyun Yang in liquid-phase rate constant calculations using \texttt{COSMO-RS}. We also thank Dr. Yunsie Chung at Merck and Professor Rudolph A. Marcus at the California Institute of Technology for insightful discussions on solvent effects on rate coefficients, and Dr. Duminda S. Ranasinghe at Montai Health for helpful discussions on excited-state calculations involving benzene and pyridine. Danica J. Adams acknowledges support provided by NASA through the NASA Hubble Fellowship Program. \copyright\space California Institute of Technology

We dedicate this paper to the memory of Professor Yuk L. Yung (1946--2026), who passed away on March 16, 2026. As one of his final scientific contributions, this work reflects his enduring vision and lasting impact on planetary science and the origins of life, and stands as a lasting testament to his legacy. We remember him with deep respect and gratitude.

\section*{Funding}\label{Funding}
J.Y. was funded by the Caltech-JPL President's and Director's Research Development Fund. D.J.A. was funded by the NASA Hubble Fellowship Program.
\section*{Supplementary Information}\label{data}
\begin{itemize}
    \item The reaction mechanism describing the thermal breakdown of a nucleobase mixture provided in \texttt{CHEMKIN} format
    \item The input file for the automatic chemical network generation using the \texttt{Reaction Mechanism Generator} (\texttt{RMG})
    \item The benzene and pyridine (both S$_0$ and T$_1$)+HCN reaction pathways imported from the gas-phase PES provided in \texttt{CHEMKIN} format
    \item The \texttt{Gaussian 09} output files of optimized geometry of all species and transition states, and intrinsic reaction coordinates analyses of transition states.
    \item \texttt{COSMO-RS} QM calculation results
    \item The outputs of the 1D-photochemical kinetic-transport atmospheric simulation of early Mars using \texttt{KINETICS}
\end{itemize} 

\section*{Author contribution}\label{contribution}
J.Y. conceptualized and led the project. J.Y. and Y.L.Y. jointly designed the study. J.Y. led the writing of the manuscript and conducted the \texttt{RMG} simulations, the \texttt{Cantera} simulations, \texttt{Gaussian 09} calculations, \texttt{COSMO-RS} calculations, and the \texttt{Arkane} rate-coefficient calculations. D.J.A. performed the \texttt{KINETICS} simulations. All authors contributed to the writing and revision of the manuscript.

\appendix
\setcounter{figure}{0}
\section{Example Appendix Section}
\label{app1}

\begin{figure*}[hbt!]
\centering
\includegraphics[width=1\textwidth]{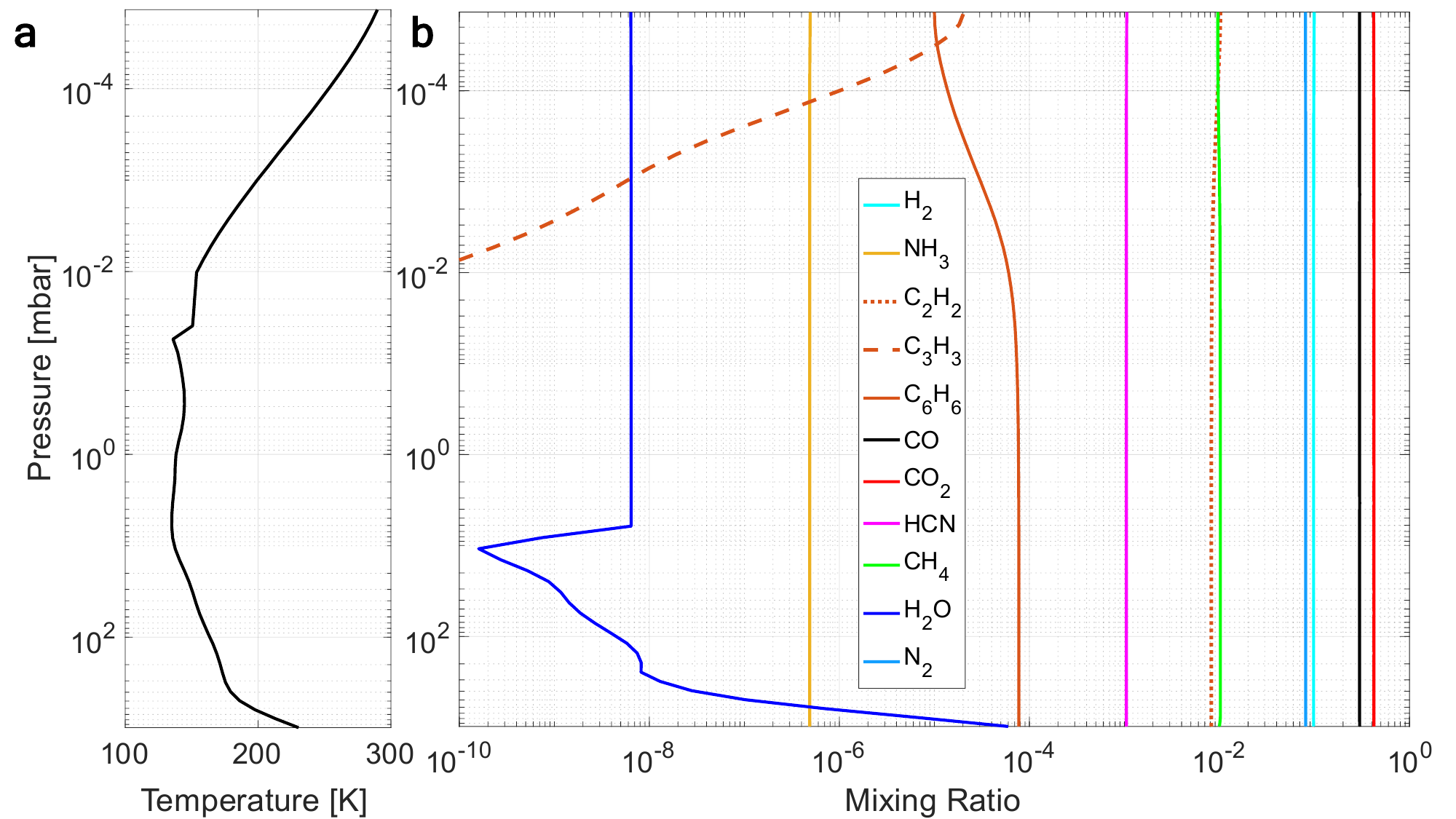}
\caption{\textbf{1D Photochemical Modeling of Early Martian Atmosphere in Cold and Dry Period.} \textbf{a}. Temperature–pressure ($T$–$P$) profile used in the 1D photochemical kinetic-transport simulation of the early Martian atmosphere, based on the cold and dry period scenario proposed by \citet{adams2025episodic}. \textbf{b}. Predicted vertical mixing ratio profiles of major atmospheric species under this scenario.}\label{fig: mars}
\end{figure*}

\begin{figure*}[hbt!]
\centering
\includegraphics[width=1\textwidth]{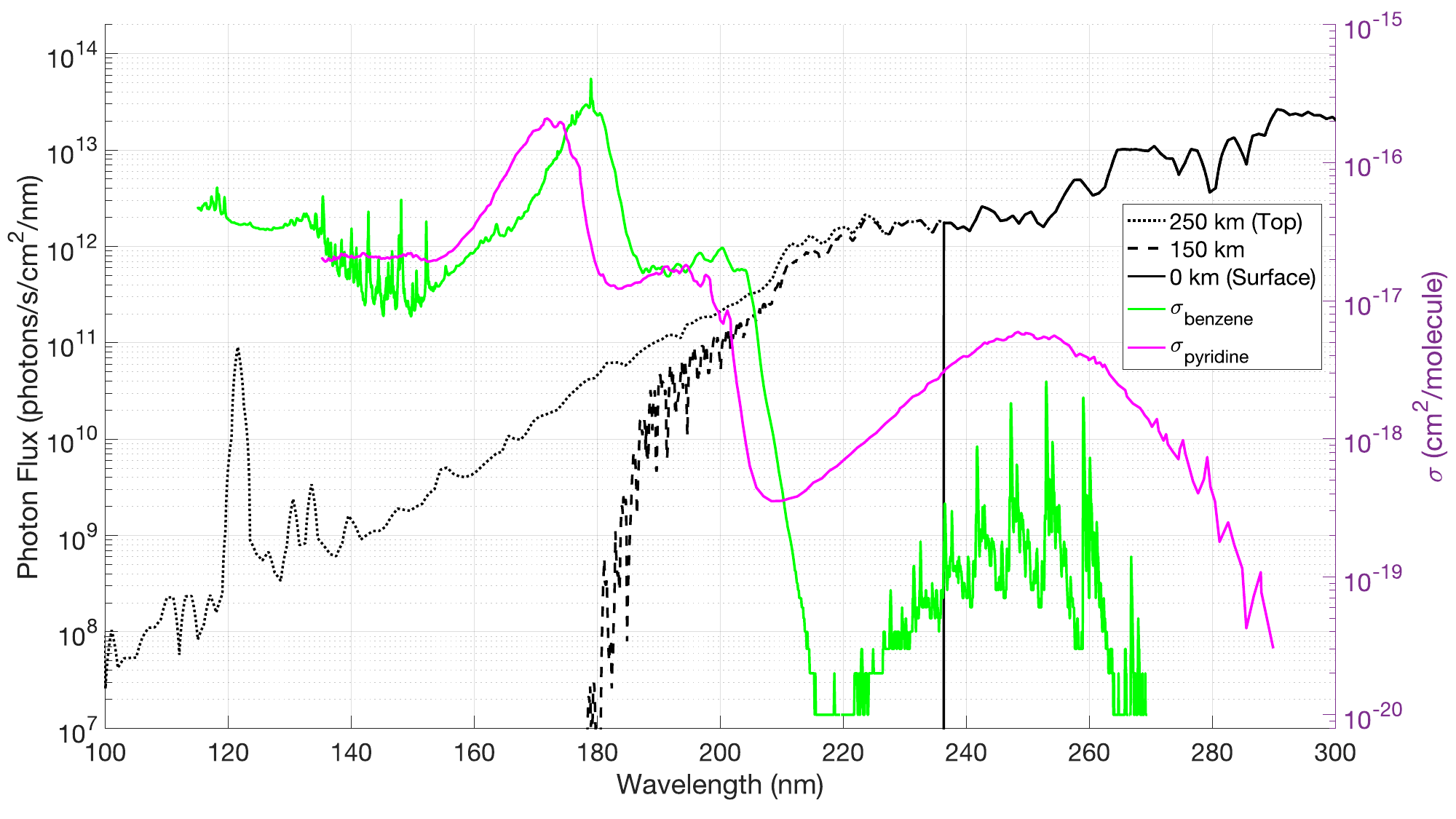}
\caption{\textbf{Wavelength vs. Photon Flux and UV-absorption cross section} The dotted black line indicates the photon flux [photons$\cdot\rm s^{-1}\cdot cm^{-2}\cdot nm^{-1}$] at each wavelength at the top of the simulated Martian atmosphere ($z$=250 km); the dashed line shows the photon flux at $z$=150 km; and the solid black line shows the photon flux at the surface ($z$=0 km). The lime line indicates the UV absorption cross section [$\rm cm^{2}\cdot molecule^{-1}$] of benzene \citep{Dawas-2017-benzene-cross-section}, while the magenta line indicates that of pyridine \citep{bolovinos1984absolute}.}\label{photons}
\end{figure*}

\bibliographystyle{elsarticle-harv} 
\bibliography{refs}



\end{document}